\documentclass[aps,prl,reprint,superscriptaddress]{revtex4-1}

\usepackage{amsmath}
\usepackage{amssymb}

\usepackage{graphicx}
\usepackage{subcaption}
\usepackage{dcolumn}
\usepackage{bm}
\usepackage{marvosym}
\usepackage{epstopdf}

\usepackage{color}
\definecolor{light-gray}{gray}{0.5}
\definecolor{blue}{rgb}{0.0,0.0,1.0}
\definecolor{green}{rgb}{0.0,0.5,0.0}
\definecolor{red}{rgb}{1.0,0.0,0.0}
\definecolor{cyan}{rgb}{0.0,0.75,0.75}
\definecolor{magenta}{rgb}{0.75,0.0,0.75}
\definecolor{yellow}{rgb}{0.75,0.75,0.0}

\newcommand{\rmc}{Rm_c^{\mathrm{turb}}}

\begin{document}
\title{The onset of turbulent rotating dynamos at the low $Pm$ limit}

\author{K. Seshasayanan}
\email{skannabiran@lps.ens.fr}
\affiliation{Laboratoire de Physique Statistique, \'Ecole Normale Sup\'erieure, CNRS, Universit\'e Pierre et Mari\'e Curie, Universit\'e Paris Diderot, 24 rue Lhomond, 75005 Paris, France}
\author{V. Dallas}
\email{v.dallas@leeds.ac.uk}
\affiliation{Laboratoire de Physique Statistique, \'Ecole Normale Sup\'erieure, CNRS, Universit\'e Pierre et Mari\'e Curie, Universit\'e Paris Diderot, 24 rue Lhomond, 75005 Paris, France}
\affiliation{Department of Applied Mathematics, University of Leeds, Leeds LS2 9JT, UK}
\author{A. Alexakis}
\email{alexakis@lps.ens.fr}
\affiliation{Laboratoire de Physique Statistique, \'Ecole Normale Sup\'erieure, CNRS, Universit\'e Pierre et Mari\'e Curie, Universit\'e Paris Diderot, 24 rue Lhomond, 75005 Paris, France}

\begin{abstract}


We demonstrate that 
the critical magnetic Reynolds number $Rm_c$
for a turbulent non-helical dynamo in the low magnetic Prandtl number $Pm$ limit (i.e. $Pm = Rm/Re \ll 1$) can be significantly reduced if the flow is submitted to global rotation.
Even for moderate rotation rates the required energy injection rate can be reduced by a factor more than $10^3$.
This strong decrease of the onset is attributed to the reduction of the turbulent fluctuations that makes the flow to have a much larger 
cut-off length-scale compared to a non-rotating flow of the same Reynolds number. 
The dynamo thus behaves as if it is driven by laminar behaviour 
(i.e. high $Pm$ behaviour) even at high values of the Reynolds number (i.e. at low values of $Pm$). Our finding thus points into a new 
paradigm for the design of new liquid metal dynamo experiments. 

\end{abstract}

\maketitle

%

The existence of planetary and stellar magnetic fields is attributed to the dynamo instability, the mechanism by which a background turbulent flow spontaneously generates a magnetic field 
by the constructive refolding of the magnetic field lines \cite{moffatt78}. 
There have been many efforts put by several experimental groups to reproduce the dynamo instability in the laboratory using liquid metals 
\cite{monchauxetal07,stieglitzmueller01,gailitisetal01,shewlathrop05,nornbergetal06,gieseckeetal12}. 
However, so far, unconstrained dynamos driven just by turbulent flows have not been achieved in the laboratory.
Successful experimental dynamos rely either in constraining the flow or using ferromagnetic materials. 
One of the major challenges to achieve a liquid metal dynamo is the large energy injection rate required 
to reach the dynamo onset that is determined by the magnetic Reynolds number $Rm = UL/\eta$
(where $U$ is the rms velocity, $L$ is the domain size and $\eta$ is the magnetic diffusivity), that should be larger than a critical value $Rm_c$.
The low value of the magnetic Prandtl number $Pm \equiv \nu/\eta \sim 10^{-5}$ of liquid metals (where $\nu$ is the viscosity),
implies that the required Reynolds number $Re = UL/\nu=Rm/Pm$ must be very large. 
Given that in turbulent flows the energy injection rate is proportional to the cubic power of $Re$ makes the dynamo onset extremely costly to reach
in the laboratory. 

From the other side, in the last decade, numerical simulations were able to reach high enough Reynolds numbers, to study the dependence of $Rm_c$ in the low $Pm$ limit
\cite{pontyetal05,mininnimontgomery05,iskakovetal07}. 
It was shown that as $Re$ was increased the turbulent fluctuations are preventing the dynamo instability resulting in a value of $Rm_c$ much larger than that of the organised laminar flows. The value of $Rm_c$ was shown to increase monotonically for values of $Pm$ around 1 but finally for high enough values of $Re$ (low enough $Pm$) a finite value of $Rm_c$ was reached independent of $Re$. This finite value is the turbulent critical magnetic Reynolds number defined as $\rmc \equiv \lim_{Re\rightarrow\infty}Rm_c$. 
Different values of $\rmc$ were obtained for the different flows under study implying that this number is not universal and that the flows can be optimised to reduce $\rmc$.
This was performed in \cite{Sadek2016} varying the forcing length scale.

In this work we propose that rotation can be used to reduce the dynamo threshold $\rmc$.
Rotation is recognized as one of the key elements that determines the main characteristics of the resulting flows and magnetic fields of planets and stars \cite{proctor94}. This is confirmed by observations over the last decade, which have measured the magnetic activity of stars as a function of their rotation period \cite{reinersetal09,morinetal10}. 
At fast rotation rates variations along the axis of rotation are suppressed rendering the flow quasi-2D in the sense that the flow 
varies weakly along the direction of rotation while retaining all three velocity components \cite{alexakis15,dt16},
a situation referred in the literature as $2.5D$ flow. These $2.5D$ flows have been shown to be effective dynamos \cite{smith2004vortex,seshasayanan2016a,seshasayanan2016b}.  
The fact that turbulent fluctuations inhibit the dynamo instability while more organised flows reduce the dynamo threshold \cite{tobiascattaneo08} 
indicates that background rotation can provide an efficient way to suppress fluctuations and
optimize the flow so that the value of $\rmc$ is reduced. 
In this Letter, we demonstrate that this is indeed the case. 
The effort to achieve the dynamo onset in rotating turbulent flows is modest in comparison 
to non-rotating turbulent flows with the columnar vortices playing a key role in the spontaneous generation of the magnetic field. 
\\

\begin{figure}[!ht]
 \begin{subfigure}{0.5\textwidth}
   \includegraphics[width=\textwidth]{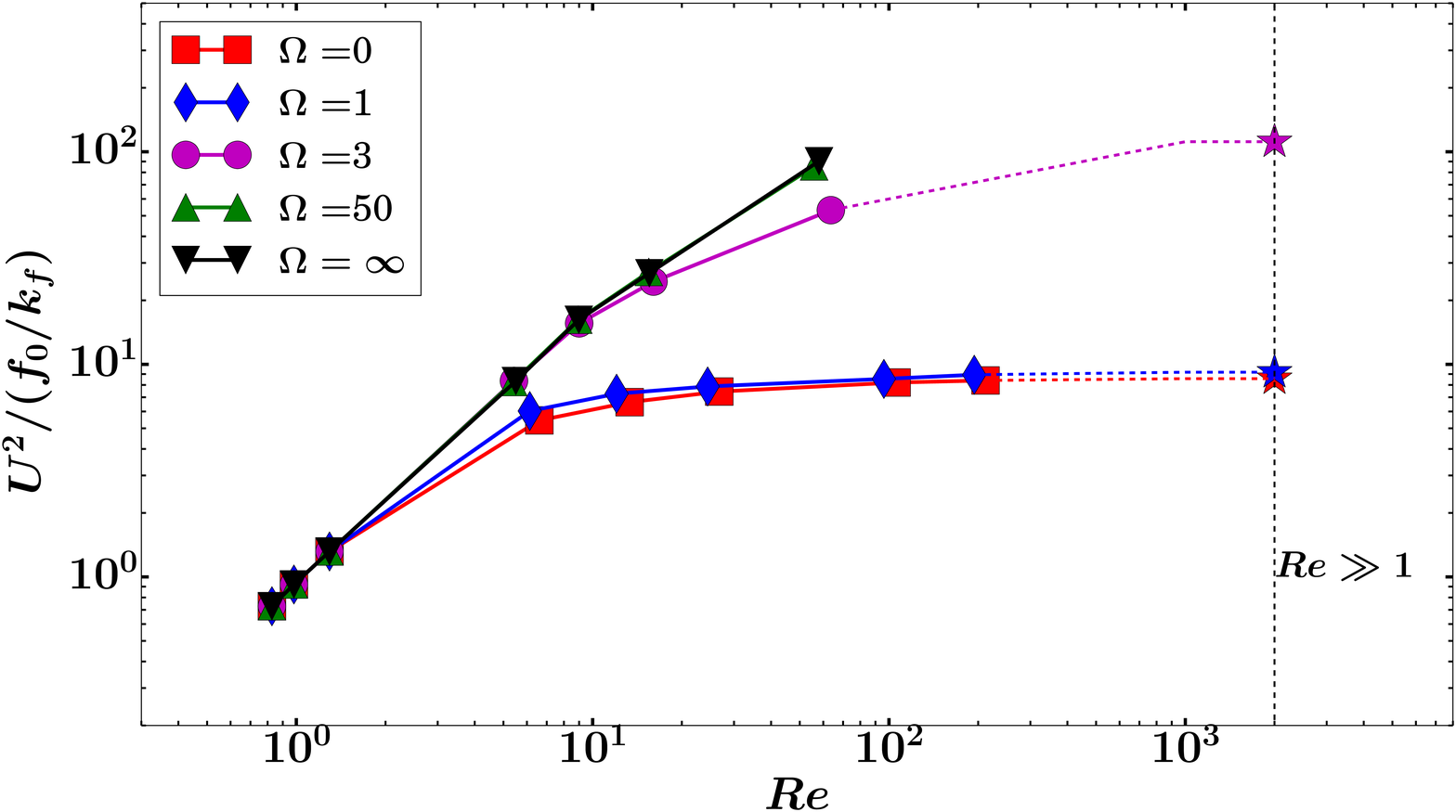}
   \caption{}
   \label{fig:U2}
 \end{subfigure}
 \begin{subfigure}{0.5\textwidth}
   \includegraphics[width=\textwidth]{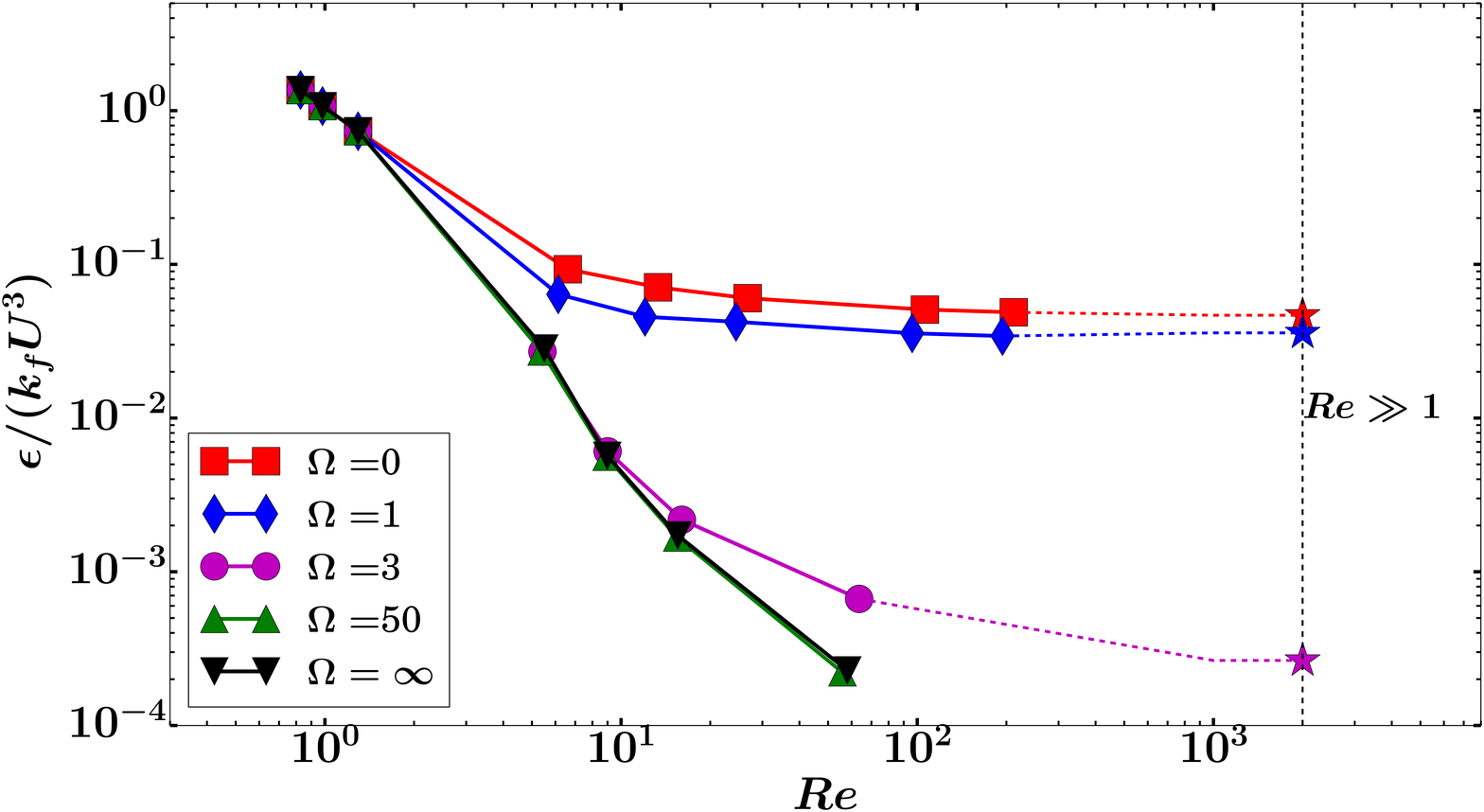}
   \caption{}
   \label{fig:eps}
 \end{subfigure}
 \caption{(Color online) The figure on top shows the normalized total velocity squared $U^2/(f_0/k_f)$ and the one on bottom shows normalized dissipations rate $\epsilon/(U^3 k_f)$ as a function of the Reynolds number $Re$ for different values of the rotation rate as mentioned in the legend. The points denoted by $\bigstar$ at $Re = 2000$ denote hyperviscosity runs. } 
 \label{fig:first} 
 \end{figure}

The governing equations involved in this study are,
\begin{align}
\partial_t {\bf u} + {\bf u} \cdot \bm{\nabla} {\bf u} = & - \frac{1}{\rho}\bm{\nabla} p - 2 \bm{ \Omega} \times {\bf u} + \nu \Delta {\bf u} + {\bf f}, \label{eqn:first}
 \\
\partial_t {\bf B} = & \bm{\nabla} \times \left( {\bf u} \times {\bf B} \right) + \eta \Delta {\bf B}.
\end{align}
where $\bf{u, B}$ are the velocity and the magnetic field respectively with $\bm{\nabla} \cdot {\bf u} =\bm{\nabla} \cdot {\bf B} = 0$, $\rho$ is the mass density, and $p$ is the pressure.
The rotation $\bm{\Omega} = \Omega \hat{\bf e}_z$ is along the $z$-direction.
We integrate these equations numerically in a cubic periodic box of length $2 \pi L$
using the pseudo-spectral code {\sc ghost} \cite{mininni2011hybrid} with a fourth-order Runge-Kutta scheme
for the time advancement and the 2/3 de-aliasing rule.
The body force  is taken to be a {\it non-helical} Roberts flow ${\bf f} = f_0 \left( \cos(k_f y), \sin(k_f x), \cos(k_f y) + \sin (k_f x) \right)$.
Since we are interested in optimizing the flow to reduce the energy consumption in dynamo experiments 
we define the non-dimensional parameters in terms of the energy injection/dissipation rate in the system measured by 
$\epsilon = \nu \left\langle |\bm{\nabla} {\bf u}|^2 \right\rangle$, where $\left\langle \cdot \right\rangle$ denotes volume and time average. 
The non-dimensional parameters in terms of $\epsilon$ are, the Reynolds number $Re = (\epsilon/k_f)^{(1/3)}/(k_f \nu)$, the magnetic Reynolds number 
$Rm = (\epsilon/k_f)^{(1/3)}/(k_f \eta)$, and the Rossby number $Ro = (\epsilon/k_f)^{(1/3)} k_f /(2 \Omega)$.
With this choice of non-dimensionalization $\rmc$
can relate directly to the power $I_c$ required to obtain dynamo by 
$I_c = \rho (2\pi L)^3k_f^4 \eta^3 \, (\rmc)^3$. 
To recover other definitions based on the rms velocity $U = \langle |{\bf u}^2| \rangle^{\frac{1}{2}} $ of the flow, $Re_{_U}=U/(k_f\nu)$ and $Ro_{_U}=U k_f/(2 \Omega)$
we provide the dependence of $\epsilon$ and $U$ on the control parameters of the system in Fig. \ref{fig:U2} and \ref{fig:eps}
and their asymptotic values in table \ref{Runs1}.
 
We are interested in different limits of the parameters in this problem. 
To model the $Re\gg1$ limit (or the $Pm \ll 1$ limit)
we also use hyperviscosity where the Laplacian in the Navier-Stokes equation (Eq. \ref{eqn:first}) is changed to $\Delta^4$. 
The other limit we would like to reach is the fast rotating limit 
$Ro \ll 1$, in which the flow becomes $2.5D$ \cite{gallet2015exact}.
The magnetic field in this case can be expressed in the form ${\bf B} = {\bf b} e^{i k_z z}$ due to the invariance of the flow along the $z$-direction. 
In this limit we follow only the $k_z = 1$ mode that was found to be the most unstable mode \cite{smith2004vortex,seshasayanan2016b}. The range of the parameters used can be found in
table \ref{Runs1}.
 
 \begin{table}[!htb]
  \caption{Numerical parameters of the simulations. 
           For all runs $f_0=1$, $L=1$ and $k_f=4$. $N$ notes the grid size.
           The reported values are for the largest values of $Re$ (regular viscosity), $\rmc$ is based on the hyperviscous runs. 
           The $\Omega = \infty$ corresponds to the $2.5D$ simulations.}
   \begin{ruledtabular}
  \centering
    \begin{tabular}{c|ccccccc} 
   $\Omega$  &   $Ro$     &    $Re$      &  $Ro_{_U}$ &   $Re_{_U}$ & N     & $\rmc$ \\ \hline
      0      &   $\infty$ &    $210$     &   $\infty$ &   $580$     & 512   & 23.6   \\ 
      1      &   $1$      &    $200$     &   $3$      &   $600$     & 512   & 34.9   \\  
      3      &   $0.21$   &    $64$      &   $2.4$    &   $720$     & 512   & 1.81   \\
     50      &   $0.011$  &    $55$      &   $0.18$   &   $920$     & 256   &  -     \\
  $\infty$   &   $0$      &    $60$      &   $0$      &   $950$     & 2048  &  -     \\
    \end{tabular}
  \end{ruledtabular}
  \label{Runs1}
\end{table}

\begin{figure}
\begin{subfigure}{0.5\textwidth}
   \includegraphics[width=\textwidth]{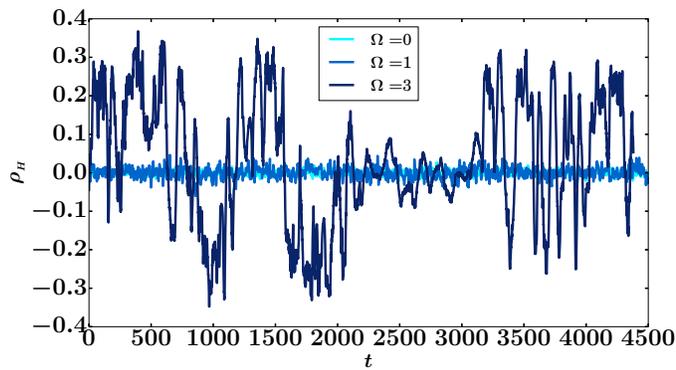}
 \end{subfigure}
  \caption{(Color online) Figure shows the relative helicity $\rho_{_H}$ as a function of time $t$ for different values of $\Omega$ mentioned in legend. Darker shades of blue correspond to larger values of $\Omega$.}
  \label{fig:hel}
 \end{figure}

We first describe the effect of rotation on the flow. 
Rotation affects the velocity field through the Coriolis term. 
At low $Re$ the flow is laminar and $\Omega$ does not modify the velocity field because the laminar flow is invariant along the direction of rotation. 
As we increase $Re$ beyond a threshold the flow becomes turbulent, varying along all three directions and hence the effect of $\Omega$ becomes more important.
%
For $\Omega \lesssim 1$, the effect of rotation is not dominant and the underlying flow is not far away from 
$3D$ isotropic turbulence. The  total 
energy $U^2$ normalised by $f_0/k_f$ and the normalized dissipation rate $\epsilon/(U^3 k_f)$
reach an asymptotic value for $Re\to \infty$ as shown in Fig. \ref{fig:first}. This asymptotic value matches with the one obtained by the hyperviscous 
simulations, which are denoted by star symbols $\bigstar$ and they are connected with the rest of the data set by dashed lines. This is the classical Kolmogorov turbulence where the large scale quantities become independent of viscosity at large $Re$.
For $\Omega =3$ the flow becomes anisotropic with lesser fluctuations along the $z$-direction. 
There is an inverse cascade present in the system that forms condensates. 
The growth of the condensate saturates when the counter-rotating vortex locally cancels the effect of global rotation
for $U\sim \Omega L$ \cite{bartello1994coherent,alexakis15}. The normalized dissipation rate $\epsilon/(U^3 k_f)$
approaches an asymptote but at a much smaller value than the non-rotating case.
For larger rotation rates $\Omega=50$ and $\Omega=\infty$ saturation comes by viscous forces and $\epsilon/(U^3 k_f)$
decreases with $Re$.
Another quantity that is important for dynamo action is the helicity $H=\left\langle {\bf u} \cdot \bm{\omega} \right\rangle$ where $\bm{\omega} = \bm{\nabla} \times {\bf u}$ is the vorticity of the flow.
Figure \ref{fig:hel} shows the normalized helicity $\rho_{_H}=H/(\|{\bf u}\| \|\bm{\omega}\|)$ as a function of time for different values of $\Omega$ (here $\|.\|$ denotes the $L_2$ norm).
As we can see for $\Omega = 3$ we observe much larger fluctuations of $\rho_{_H}$ whose average over time is zero.
Note that the time scale of the fluctuations is much longer than the eddy turnover timescale $L/U\simeq 0.2$.
These fluctuations are due to the formation of the condensate \cite{dt16}. 
At small $\Omega$ the helicity fluctuations are governed by the small scales for which the eddy turnover time is very small, for large $\Omega$ the helicity fluctuation is governed 
by the $k_z = 0$ mode which fluctuates over a much larger time scale. A priori we do not know whether 
the transition from a flow with no inverse cascade to a flow with an 
inverse cascade will decrease the dynamo threshold.
 
\begin{figure}[!ht]
   \includegraphics[width=0.5\textwidth]{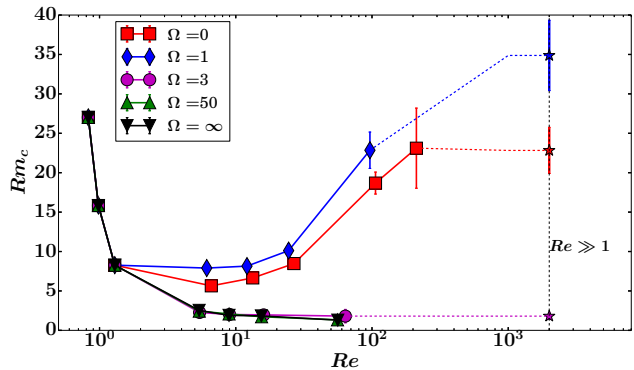}
   \caption{(Color online) Figure shows the critical magnetic Reynolds number $Rm_c$ as a function of $Re$ for different values of $\Omega$ shown in the legend. The points denoted by $\bigstar$ at $Re = 2000$ denote hyperviscosity runs.}
  \label{fig:rmc}
 \end{figure}

Now we look at the effect of rotation on dynamo. 
Figure \ref{fig:rmc} shows the critical Reynolds number $Rm_c$ as a function of $Re$ for different values of $\Omega$.
%
To calculate $Rm_c$ we run simulations of the same flow (same $Re$ and $Ro$) 
but with different values of $Rm$. 
$Rm_c$ was determined by linear interpolation of the growth-rate
between dynamo (positive growth rate) and no-dynamo runs (negative
growth rate).
%
The cases of $\Omega = 0$ and 1 display similar behaviour as other
 studies of non-rotating flows (see \cite{pontyetal05,mininnimontgomery05,iskakovetal07}) 
 in which $Rm_c$ initially increases with $Re$, until it begins to become constant at large $Re$.
 For $\Omega=1$ the asymptotic value $\rmc$ is larger than the
 $\Omega=0$ case expressing an initial hindering effect for the dynamo by rotation.
 For $\Omega \geq 3$ however we see a much lower threshold for the dynamo
 instability and no such increase due to turbulence is observed. In fact the threshold
 for $\Omega=3$ does not appear different from the $\Omega =\infty $
 implying that the destructive effect of the 3D turbulent fluctuations on
 dynamo has already disappeared. The ratio between the $\rmc$ for the case of $\Omega = 0$ and the case
 $\Omega = 3$ for the hyperviscous runs is approximately $\sim 12$. The
 injected power $\epsilon$ scales like $Re^3$ implying a reduction in the
 power required for a dynamo instability 
by a factor of $2\cdot10^3$ between $\Omega=0$ and $3$ and a factor of $8\cdot10^3$ between
$\Omega=1$ and $3$ (see Fig. \ref{fig:rmc}).
%
%

\begin{figure}[!ht]
 \begin{subfigure}{0.5\textwidth}
   \includegraphics[width=\textwidth]{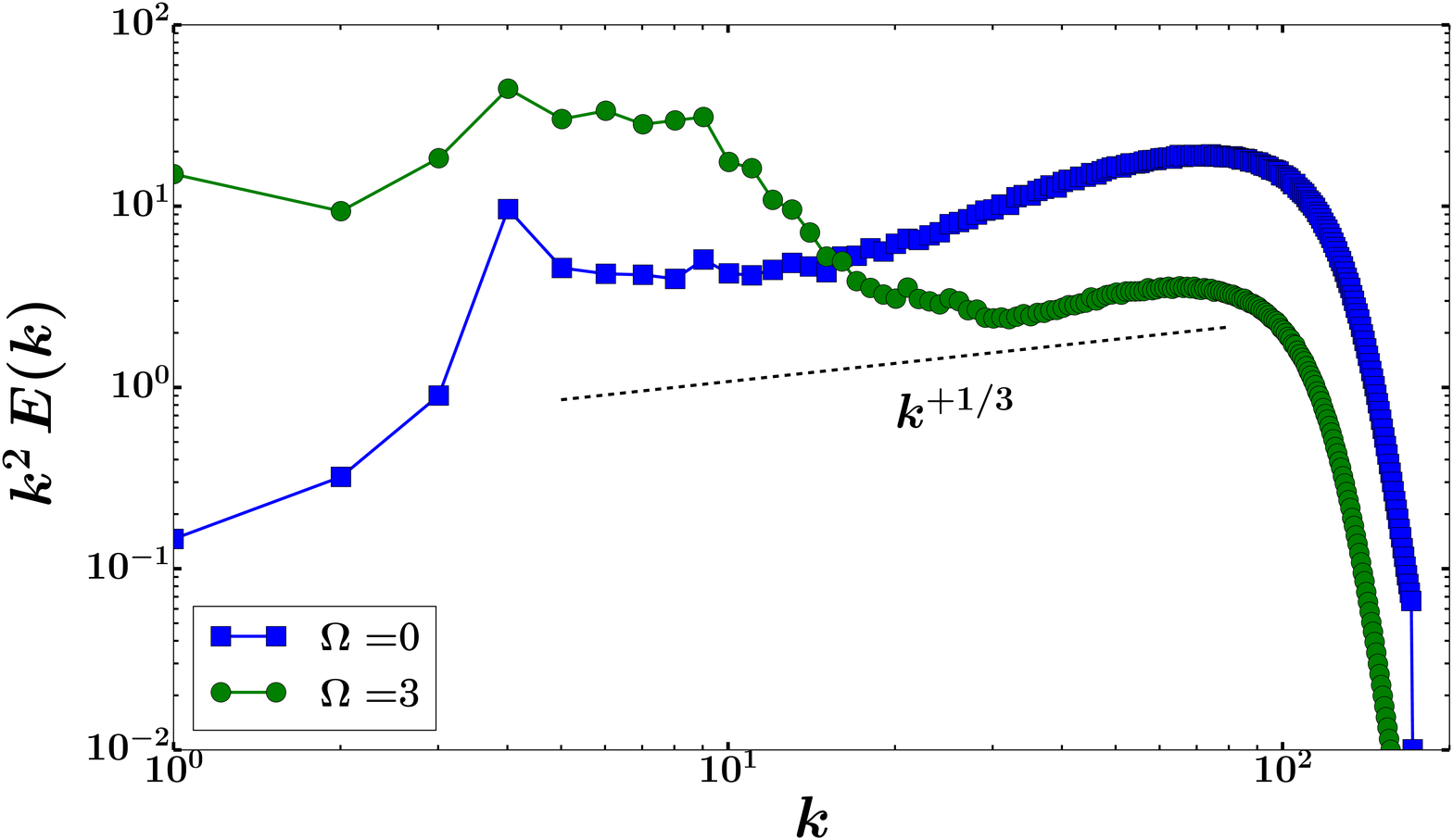}
   \caption{}
   \label{fig:kspec}
 \end{subfigure}
 \begin{subfigure}{0.5\textwidth}
   \includegraphics[width=\textwidth]{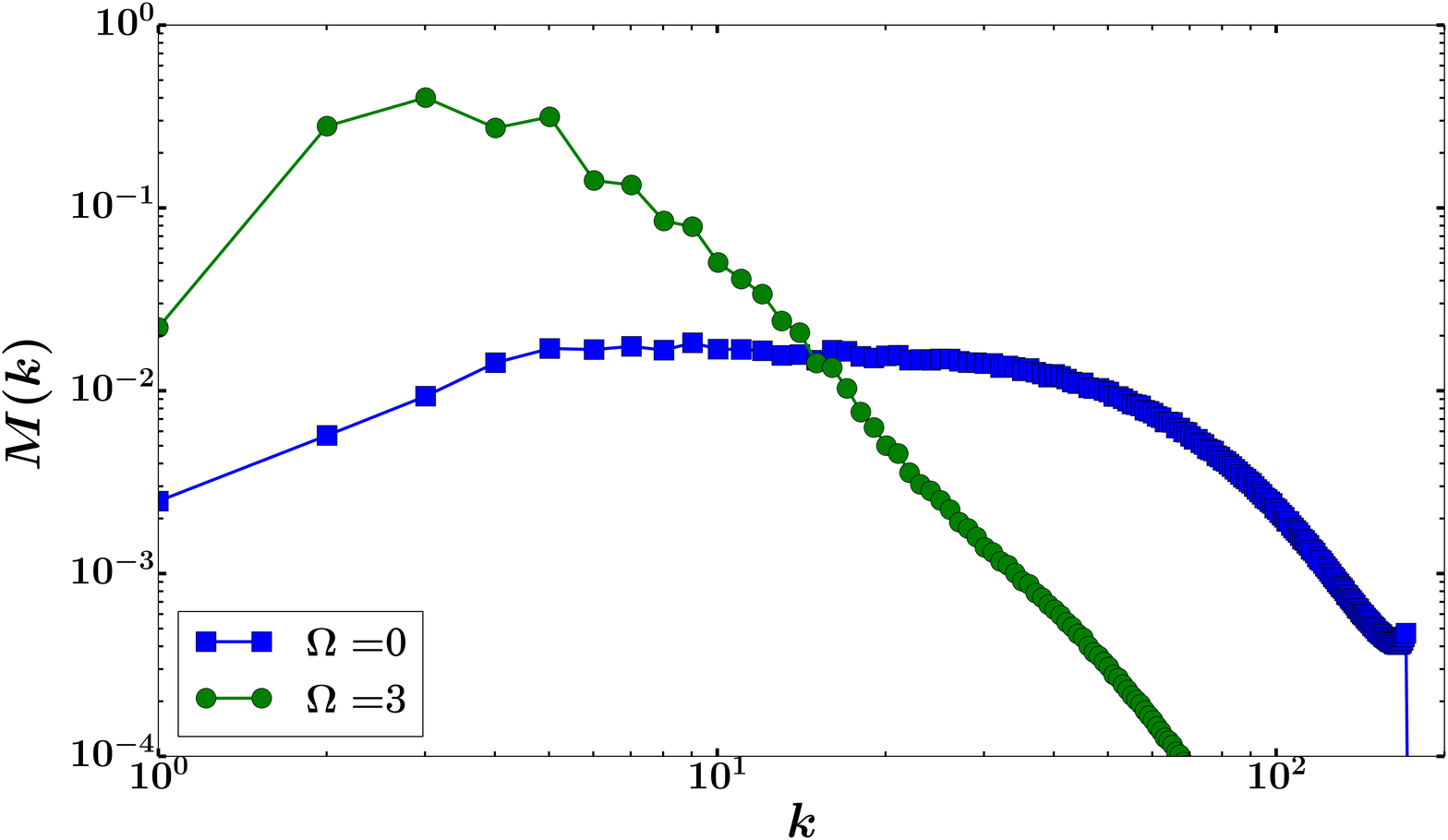}
   \caption{}
   \label{fig:mspec}
 \end{subfigure}
  \caption{(Color online) Figure shows in a) the compensated kinetic energy spectra $k^2 E(k)$ in b) the magnetic energy spectra for the two different cases of $\Omega = 0,3$ for the hyperviscous runs. }
  \label{fig:spectra}
 \end{figure}

\begin{figure}[!ht]
   \includegraphics[width=0.4\textwidth]{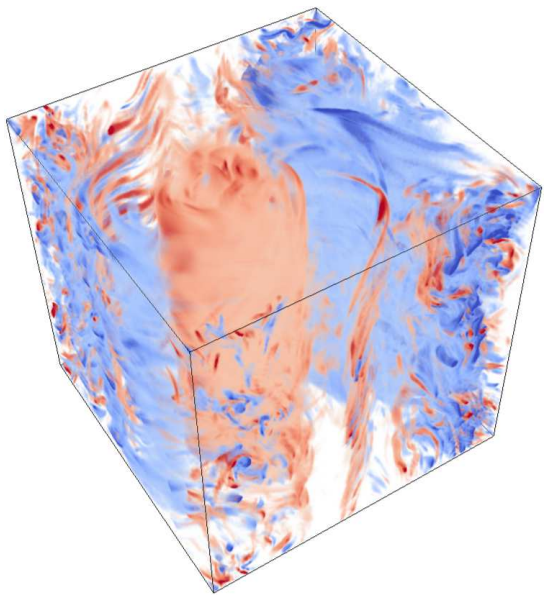}
   \caption{(Color Online) Figure shows the vertical vorticity ${\bm \omega}_z$ as a function of space for the parameter $\Omega = 3, Re \approx 60$.
   Red corresponds to positive (co-rotating) values and blue to negative (counter-rotating) values.  }
  \label{fig:omega_vizual}
 \end{figure}

To decipher the reason behind this drop in $\rmc$ at $\Omega = 3$ we display in Fig. \ref{fig:kspec} the enstrophy spectra $k^2E(k)$
for $\Omega = 0$ and $ \Omega = 3$ obtained from the hyper-viscous runs. 
Large enstrophy implies a larger stretching rate of the magnetic field lines (although not necessarily constructive).
For $\Omega=0$ a close to Kolmogorov behavior is observed with the enstrophy spectrum $k^2E(k)$ increasing with $k$ 
after the forcing scale $k_f=4$. The strongest stretching rate is thus clearly at the small incoherent scales. 
On the contrary for $\Omega=3$ the enstrophy spectrum $k^2E(k)$ is decreasing with $k$. 
Only at the smallest scales $k^2E(k)$ starts to increase again.  
Thus, the small scale fluctuations are suppressed and the dominant stretching rate $u_{\ell}/\ell$ is restricted to the large coherent scales.  
Figure \ref{fig:omega_vizual} shows the vertical vorticity field ${\bm \omega}_z$ for $\Omega = 3$
displaying a strong coherent co-rotating vortex aligned with the global rotation and 
a counter rotating vortex responsible for the energy cascade to small scales \cite{alexakis15}.
The dynamo thus behaves as if it is driven by an organised laminar flow 
(i.e. high $Pm$ behaviour) even at very large Re (i.e. at low values of Pm).
We note that this suppression of small scale fluctuations is not due to a dissipative mechanism since the Coriolis term is not dissipative
and thus does not lead to an extra cost in energy injection.  

The magnetic energy spectra for $Rm$ close to the onset are shown in Fig. \ref{fig:mspec}. 
 \begin{figure}[!ht]
   \includegraphics[width=0.4\textwidth]{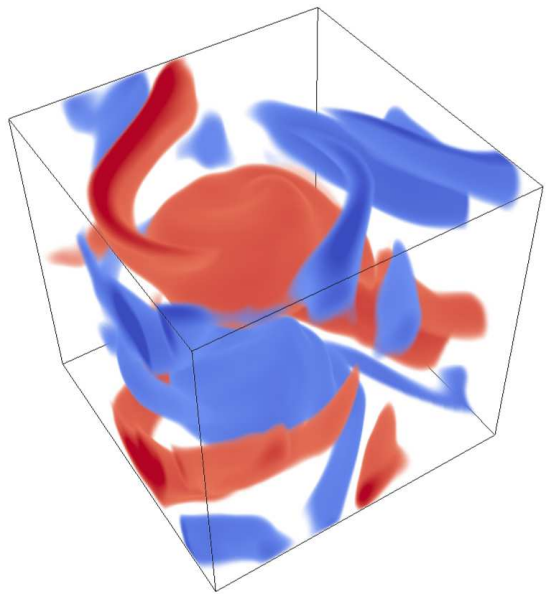}
   \caption{(Color Online) Figure shows the vertical current $j_z$ as a function of space for the parameter $\Omega = 3, Re \approx 60$.
   Red corresponds to positive and blue to negative values.}
  \label{fig:vizual}
 \end{figure}
For the case of $\Omega = 0$ the magnetic energy spectrum 
is almost flat with an exponential decay at high wavenumbers.
The unstable eigenmode (not shown here) takes the form of thin filamentary structures.
On the other hand, for $\Omega = 3$ the magnetic energy spectrum decreases fast with $k$
and its relative dissipation rate is thus not as strong. The structure of the vertical current field $\bm j_z$ from an unstable
eigenmode of the dynamo at $\Omega = 3$, is shown in Fig.\ref{fig:vizual}.  
The magnetic field as seen previously in the spectra is present at large scales,
with the $k_z=1$ being dominant. Most of the magnetic energy is concentrated 
along the coherent co-rotating vortex in 
two counter directed spiral flux tubes. 
 

The present study shows that global rotation can play a positive role in the dynamo instability by suppressing turbulent fluctuations. 
This non-dissipative suppression leads the flow to drive the dynamo by well organized large scales that have long correlation times and
thus are more effective in performing a constructive refolding of the magnetic field lines. 

This discovery provides a new paradigm for the design of new dynamo experiments that include global rotation. 
Reaching rotation rates in the laboratory that lead to quasi-2D flows 
is feasible and has been achieved in water-tank experiments \cite{campagne2014direct,yarom2013experimental}. 
The additional energy cost for maintaining the rotation is probably minimal compared to the large gain 
of the order of $10^3$  due to the suppression of turbulent fluctuations.
The only issue that needs to be considered is that the design of the domain and
the forcing should guarantee that all three velocity components are present, so that the flow becomes $2.5D$ and not $2D$.
Finally, we note that this result also shows that in fast rotating systems, 
like the Earth, the critical magnetic Reynolds number based on the injected energy to sustain a dynamo instability might stay very small 
$Rm \sim O(1)$ even at large Reynolds numbers.



%
\begin{acknowledgements}
The authors acknowledge enlightening discussions with S. Fauve. V.D. acknowledges support from the Royal Society and the British Academy of Sciences (Newton International Fellowship, NF140631). The computations were performed using the HPC resources from GENCI-TGCC-CURIE (Project No.x2016056421) and ARC1, part of the High Performance Computing facilities at the University of Leeds, UK.
\end{acknowledgements}

\bibliography{references}
\end{document}